\documentclass[journal=jacsat]{achemso}

\usepackage[utf8]{inputenc}
\usepackage{textgreek}
\usepackage[version=3]{mhchem} % Formula subscripts using \ce{}
\usepackage{chemformula} % Formula subscripts using \ch{}
\usepackage[utf8]{inputenc}
\usepackage[T1]{fontenc} % Use modern font encodings
\usepackage{booktabs} % For \toprule, \midrule, \bottomrule
\usepackage{textgreek}
\author{Wenxuan Cai}
\affiliation[Imperial College London]
{Department of Materials, Imperial College London, London, UK}

\author{Stefan Kucharski}
\affiliation[University College London]
{Department of Chemistry, University College London, London, UK}

\author{Chris Blackman}
\affiliation[University College London]
{Department of Chemistry, University College London, London, UK}

\author{Juhan Matthias Kahk}
\affiliation[University of Tartu]
{Institute of Physics, University of Tartu, Tartu, Estonia}

\author{Johannes Lischner}
\email{j.lischner@imperial.ac.uk}
\affiliation[Imperial College London]
{Department of Materials, Imperial College London, London, UK}

\title[An \textsf{achemso} demo]
  {Predicting core-level X-ray photoemission spectra of oxide surfaces from first principles - a case study for SnO$_2$}

\abbreviations{X-ray photoemission spectroscopy, tin dioxide, surface chemistry}

\begin{document}

\begin{tocentry}

\begin{figure}[H]
    \centering
    \includegraphics[width=0.9\textwidth]{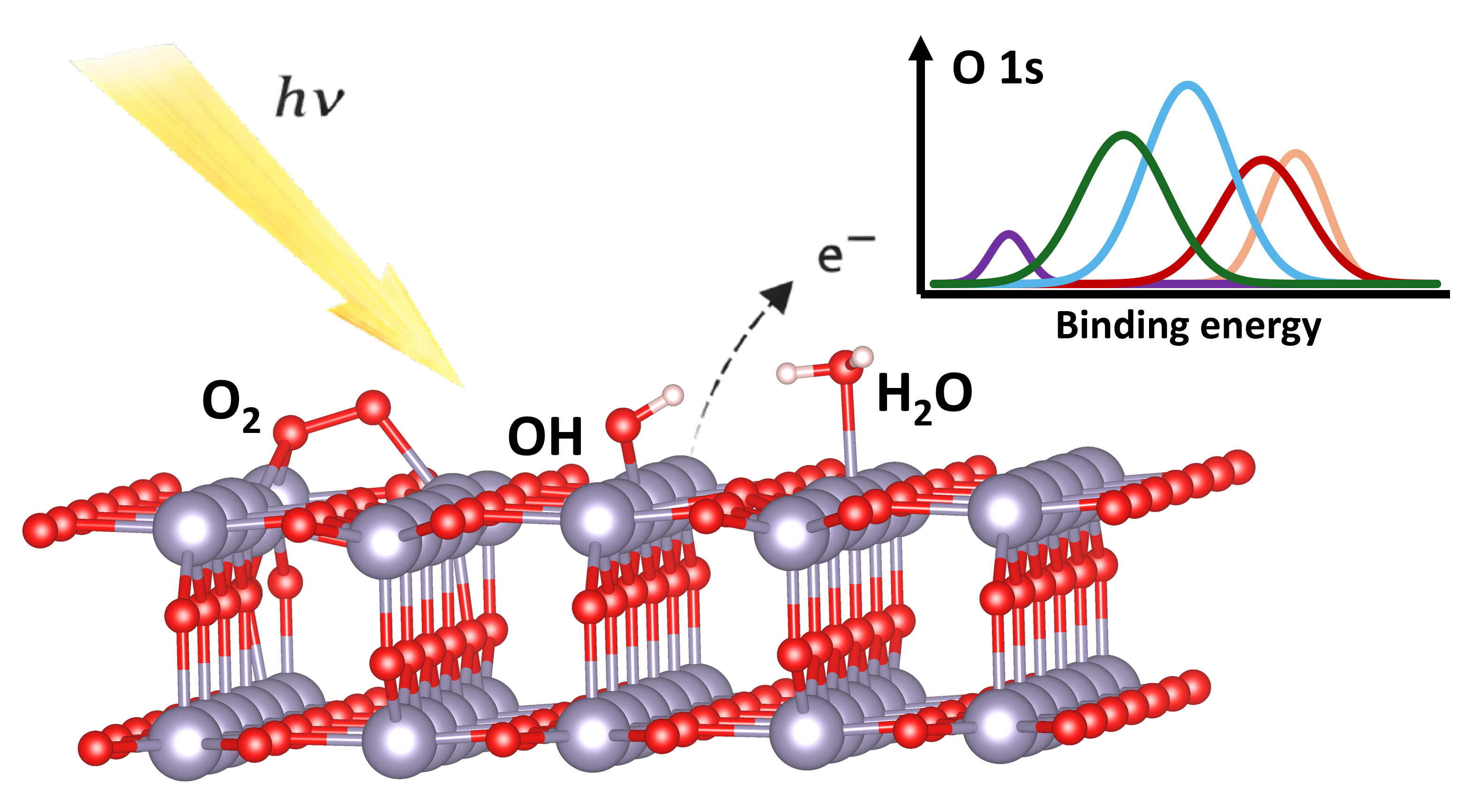}
\end{figure}

\end{tocentry}

\begin{abstract}

X-ray photoemission spectroscopy (XPS) is a powerful technique to gain insight into the chemical properties of oxide surfaces. However, the interpretation of XPS spectra is notoriously difficult as realistic surfaces contain different terminations, reconstructions, adsorbates and defects all of which leave (potentially overlapping) spectroscopic fingerprints. To address this challenge, we present a first-principles approach based on the Z+1 method that allows us to predict XPS spectra of oxide surfaces which can directly be compared to experimental measurements. We present results for different SnO$_2$ (110) surfaces: the stoichiometric surface, surfaces with different types of vacancies (one of which is the fully reduced surface) and also the fully reduced surface with adsorbed OH and O$_2$ molecules. For these systems, we calculate the O 1s core-electron binding energies of all oxygen atoms and then use this to predict the XPS spectrum. We find that the fully reduced surface gives rise to a highly symmetric peak shape in agreement with recent XPS measurements. In contrast, the spectrum of the stoichiometric surface exhibits an additional feature at low binding energies caused by the bridging oxygen atoms at the surface. For the reduced surface with OH and O$_2$ adsorbates, the spectrum exhibits additional features at higher binding energies. The predicted spectra are in good agreement with experimental results obtained for reduced surfaces that have been exposed to oxygen gas. The presented method is general and can be straightforwardly applied to other surfaces.   
  
\end{abstract}

\section{Introduction}

Tin dioxide (SnO$_2$) is an \textit{n}-type semiconductor which is used in transparent conducting electrodes ~\cite{jiang2012,ginley2000,minami2000,fortunato2007,kiruthiga2022}, catalytic systems~\cite{wachs2005,hassan2018,zhang2024,liu2015}, batteries~\cite{chen2013,chen2012,park2007}, corrosion protection~\cite{subasri2003,osterwald2003}, and as a conductometric gas sensor~\cite{chavali2019,sun2012,wang2010,gopel1995,das2014,barsan1999}. For many of these applications, surface chemical processes play a key role. However, a complete model of SnO$_2$’s surface chemistry under different conditions is currently still lacking. For example, the microscopic mechanism that gives rise to the change of resistance upon exposure to oxygen remains unclear ~\cite{das2014,gopel1995}. Traditionally, it was believed that this phenomenon could be explained by ionosorption theory which states that oxygen molecules adsorb on the surface, accept electrons and then dissociate to form negatively charged adatoms~\cite{gurlo2006,KOROTCENKOV2017,chris2021}. More recently, an alternative mechanism based on the concentration change of charged oxygen vacancies upon oxygen gas exposure has been put forward~\cite{MAIER1988,reghu2018,chris2021}. To decide which theory is correct, various experimental techniques for measuring surface chemistry have been applied to SnO$_2$, including electron paramagnetic resonance spectroscopy~\cite{NAVALE2008,PARTHIBAVARMAN2013,li2023}, temperature programmed desorption~\cite{ega1981,Yoshinobu1990,ZHANG2015} and X-ray photoemission spectroscopy (XPS)~\cite{KWOKA200536,BABU2018211,kucharski2021,kucharski2022direct,kucharski2025}.

XPS provides insights into the chemical structure of materials by measuring the binding energy of core electrons ~\cite{kucharski2021,kucharski2025,KWOKA200536,AKGUL201350,BABU2018211}. For example, Kucharski and coworkers performed an in-operando XPS study of SnO$_2$ gas sensors at near ambient oxygen pressure conditions and were able to correlate the measured conductivities with observed XPS spectra~\cite{kucharski2021,kucharski2025}. In particular, they observed significant changes in the measured O 1s peak height upon oxygen exposure which they interpreted as a change of the O/Sn ratio caused by the “healing” of vacancies by dissociated oxygen molecules. Moreover, they observed an additional peak in the O 1s spectrum at higher binding energies than the main peak upon exposure to O$_2$ gas. They interpreted this as an additional oxygen chemical environment and hypothesized that this could be an undissociated oxygen molecule adsorbed to a bridging oxygen vacancy. However, assigning peaks in measured XPS spectra to specific chemical environments in a sample is notoriously difficult as accurate reference spectra for the many potential adsorbate configurations are often not available (or available reference spectra for the same chemical environment disagree with each other)~\cite{tabata2003,KWOKA200536,chuvenkova2015,wang2022}.

First-principles materials modelling techniques can provide valuable insights into possible surface structures. For example, Oviedo and Gillan used \textit{ab initio} density functional theory (DFT) to investigate the energetics of oxygen adsorption on stoichiometric and also weakly and strongly reduced SnO$_2$ (110) surfaces~\cite{OVIEDO2000,OVIEDO2001,OVIEDO2002}. They reported strong adsorption of molecular oxygen at bridging oxygen vacancies and weaker adsorption at the fivefold coordinated Sn atom. To model an \textit{n}-doped surface, Sopiha and coworkers performed DFT calculations of SnO$_2$ surfaces with additional doped electrons and reported the formation of superoxide O$_2^-$ species on (110) and (101) surfaces as well as doubly ionized O$^{2-}$ species on (100) facets~\cite{kos2021,shaban2025unraveling}. To assess whether these predicted surface structures are consistent with XPS experiments, it is necessary to calculate core-electron binding energies of the surface structures. In recent years, significant advances have been made in this direction. For example, Kahk and Lischner demonstrated that the $\Delta$-Self-Consistent-Field ($\Delta$SCF) approach yields highly accurate absolute core-electron binding energies for molecules, bulk metals and also metallic surfaces and Kahk and coworkers later extended the approach to non-metallic materials ~\cite{kahk2018,kahk2019,kahk2021,kahk2022,kahk2023}. However, only very few computational studies of core-electron binding energies have been performed for oxide surfaces with adsorbed species~\cite{LiNolan2023,fongkaew2017core}.

In this paper, we calculate O 1s core-electron binding energy (CEBE) shifts and the corresponding XPS spectra of different SnO$_2$ surfaces using the Z+1 approach in which the nucleus of the atom with the core hole is replaced by a nucleus with an additional proton. Results are presented for the stoichiometric (110) surface, (110) surfaces with different types of oxygen vacancies and also the fully reduced (110) surface with adsorbed OH and O$_2$ groups. CEBE shifts obtained from the Z+1 approach are in excellent agreement with those obtained from the $\Delta$-SCF approach. We find that bridging O atoms of the stoichiometric surface give rise to an additional peak at lower binding energies, while both OH and O$_2$ adsorbates on the fully reduced surface produce peaks at higher binding energies than the main peak. The O 1s spectrum of the fully reduced surface is found to be highly symmetrical. The calculated spectra are in excellent agreement with the XPS measurements of Kucharski and coworkers~\cite{kucharski2021,kucharski2025} providing microscopic insights into the chemical nature of the oxygen environments responsible for the observed XPS signals. Our approach is general and can be applied to many other interesting surface chemistry problems in the future.

\section{Methods}

In an XPS measurement, a photon is used to remove an electron from the core state of an atom in the sample creating a core hole. To model the atom with the core hole, we employ the Z+1 approximation, i.e. the core hole and the nucleus of the atom are replaced by a nucleus with an additional proton. For example, for an O atom with a 1s core hole, instead of removing a 1s electron, the nuclear charge is increased by one proton charge, i.e. the F nuclear is used. This avoids the need to perform calculations with electronic occupancies that do not follow the Aufbau principle, which can be difficult to converge~\cite{BAGUS1999}. It is important to note, however, that only accurate binding energy shifts, i.e. differences between binding energies, can be obtained from this method. For the surface calculations presented in this work this is not a significant limitation as a useful reference is provided by the CEBE of a "bulk" atom, i.e. an atom far below the surface. To assess the accuracy of the Z+1 approximation, we also perform calculations with non-Aufbau principle occupancies using the full $\Delta$-SCF approach.

\begin{figure}[H]
    \centering
    \includegraphics[width=0.95\textwidth]{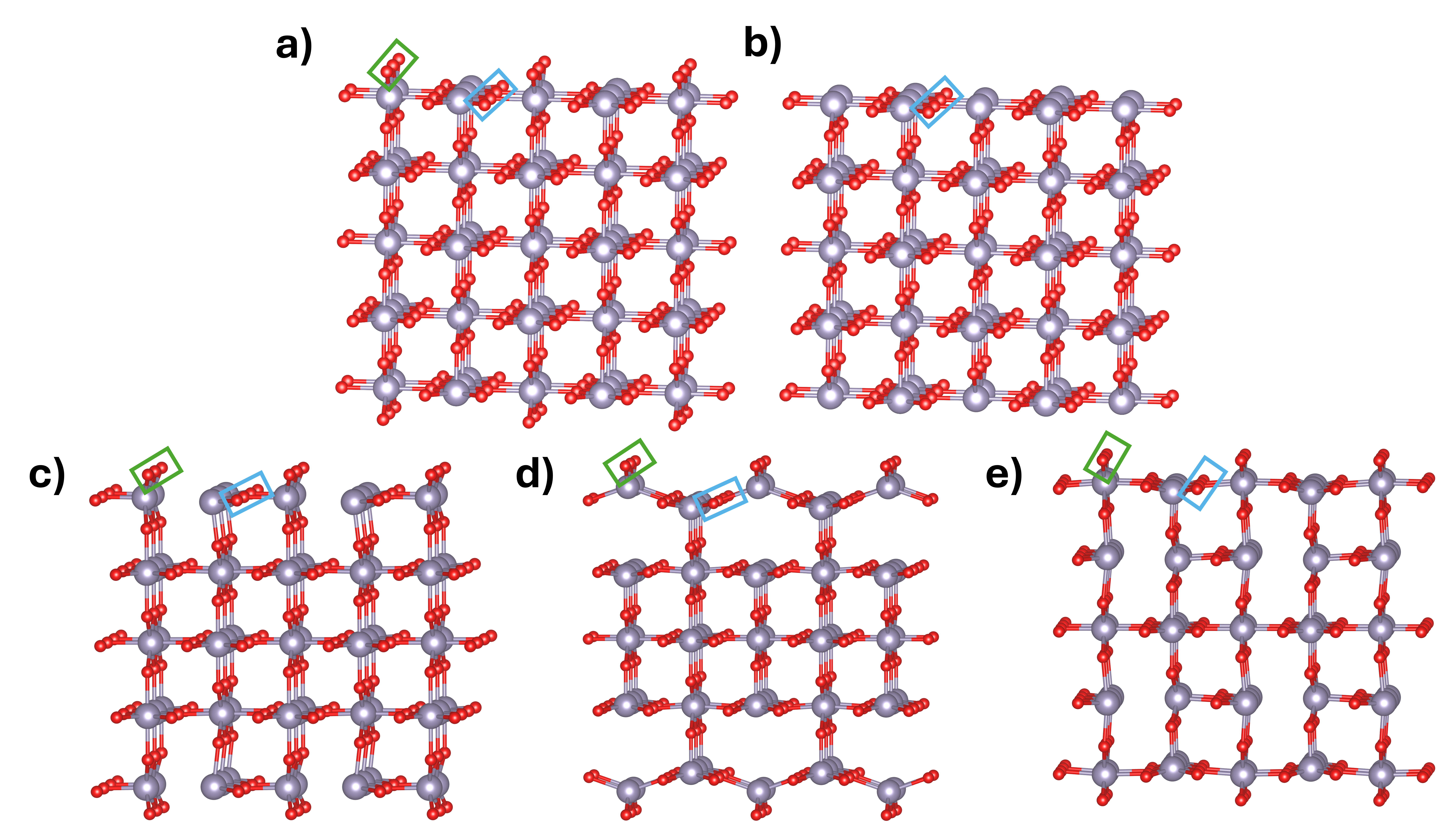}
    \caption{Atomic structure of the initial SnO$_2$ (110) surfaces studied in this work: (a) stoichiometric surface, (b) fully reduced surface, (c) stoichiometric surface with in-plane vacancy, (d) stoichiometric surface with sub-bridging vacancy, and (e) stoichiometric surface with sub-in-plane vacancy. Tin atoms are shown in grey and oxygen atoms in red. Bridging oxygen atoms and in-plane oxygen atoms are indicated by green and blue boxes, respectively.}
    \label{fig:slab_model_UHV}
\end{figure}

\begin{figure}[H]
    \centering
    \includegraphics[width=0.95\textwidth]{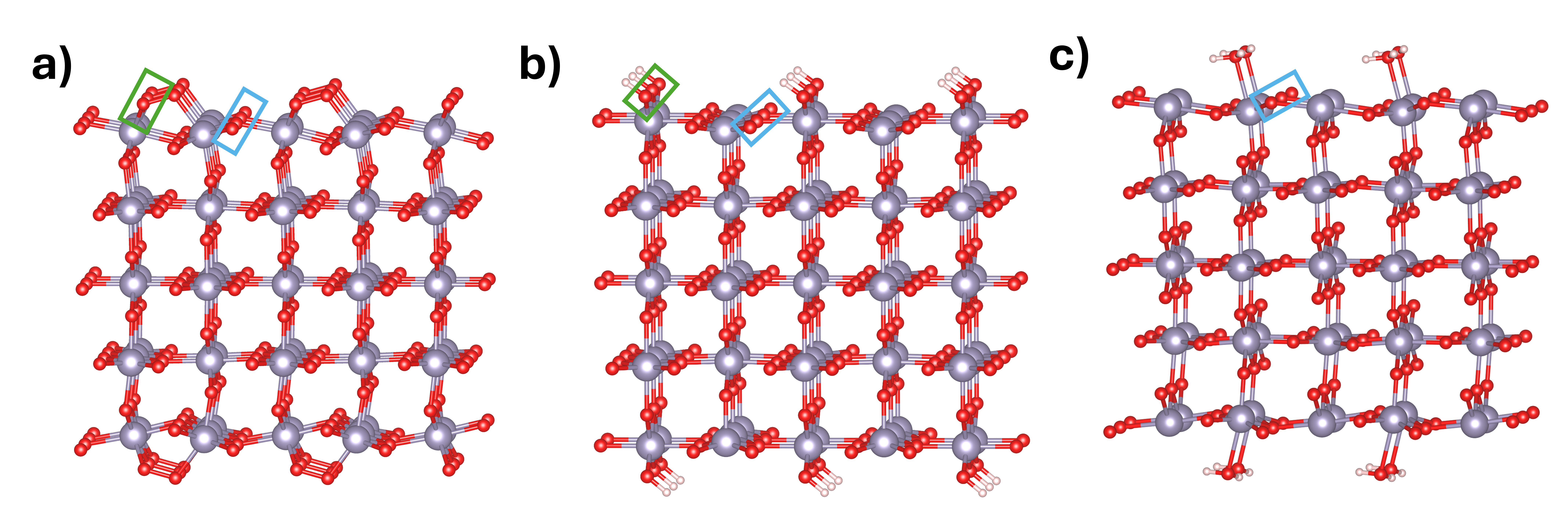}
    \caption{Atomic structure of the adsorption SnO$_2$ (110) surfaces studied in this work: (a) fully reduced surface with adsorbed O$_2$ molecules, (b) fully reduced surface with an adsorbed OH, and (c) fully reduced surface with adsorbed H$_2$O molecules. Tin atoms are shown in grey and oxygen atoms in red. Bridging oxygen atoms and in-plane oxygen atoms are indicated by green and blue boxes, respectively.}
    \label{fig:slab_model_O2}
\end{figure}

To model CEBEs of SnO$_2$ surfaces, we employ a periodic slab model of the surface. In particular, we model a rutile (110) surface with a thickness of up to 7 layers. We study surfaces both with and without adsorbates. Fig.~\ref{fig:slab_model_UHV} shows the atomistic structures of the adsorbate-free surfaces which include the stoichiometric surface (panel a)) as well as surfaces with different types of oxygen vacancies: the surface with all bridging O atoms removed (which we will refer to as the fully reduced surface, see panel b)), the surface with in-plane O vacancies (panel c)), with sub-bridging O vacancies (panel d)) and sub-in-plane O vacancies (panel e)). Fig.~\ref{fig:slab_model_O2} shows the surfaces with adsorbates which include the fully reduced surface with adsorbed O$_2$ molecules (panel a)), with OH groups (panel b)) and with H$_2$O molecules (panel c)). All structures have been relaxed until the maximum residual force component for all atoms is less than $10^{-5}$~eV/\AA.

When a core hole is introduced (corresponding to the replacement of an O atom with an F atom in the Z+1 approximation), large supercells are required in order to avoid spurious interactions between periodic images of the core hole. The remaining effect of such interactions can be removed using an electrostatic continuum model following the work of Freysoldt and coworkers, which was originally developed for charged surface defects~\cite{freysoldt2018}.

Based on the calculated CEBE shifts of all O atoms at the surface, we construct the simulated O 1s XPS spectrum $I(E)$ as a weighted sum of Gaussians, where the weights are determined by the probability of the photoelectron escaping from a certain depth below the surface, according to
\begin{equation}
I(E) \propto \sum_i e^{-z_i/\lambda} e^{-(E - E_i)^2/(2\sigma^2)}.
\end{equation}

Here $E_i$ denotes the CEBE of O atom $i$ with z-coordinate $z_i$ relative to the z-coordinate of the in-plane O atoms. Also, $\lambda$ is the mean free path of electrons and $\sigma$ captures the various broadening mechanisms. Here, we use $\lambda = 2$~\AA \; and $\sigma = 0.9$~eV so that the calculated spectra closely resemble the measured spectra obtained by Kucharski and coworkers~\cite{kucharski2021}. Further work is needed to understand why such a small value of $\lambda$ is required to obtain agreement with experiment. For the evaluation of the XPS spectrum, we have set the z-coordinates of the bridging O atoms and also of the adsorbate O atoms to zero as photoelectrons emitted from these atoms are unlikely to undergo any scattering events. 

All calculations were performed using the all-electron FHI-aims code, which employs numeric atom-centered orbitals~\cite{blum2009}. The scaled zeroth-order regular approximation (ZORA) is used to capture relativistic effects~\cite{zora1994}. For the relaxations and the Z+1 calculations, we employ the Perdew-Burke-Ernzerhof (PBE) exchange-correlation functional~\cite{pbe1996} as well as a "tight" setting for the basis set. To calculate the density of states, we use the B3LYP functional~\cite{lee1988,becke1993}. For the unit cell of the stoichiometric SnO$_2$ surface, we used a $4\times6\times1$ $k$-point grid, which was reduced in supercell calculations to ensure a consistent sampling of k-space. A vacuum layer of at least 50~\AA{} was used to separate periodically repeated slabs.

\section{Results and discussion}

\begin{table}[H]
\centering
\caption{Comparison of O~1s CEBE shifts from $\Delta$-SCF and Z+1 methods for bridging oxygen atoms and in-plane oxygen atoms of a stoichiometric SnO$_2$(110) surface. Calculations were performed using $3 \times 3$ supercells for both 3-layer and 5-layer slabs. All shifts are referenced to the O atom in the center of the slab.}
\begin{tabular}{cccccc}
\toprule
Site & No. of layers & $\Delta$-SCF shift (eV) & Z+1 shift (eV) & Difference (eV) \\
\midrule
Bridging O & 3 & 1.48 & 1.46 & 0.02 \\
Bridging O & 5  & 1.49 & 1.47 & 0.02 \\
In-plane O & 3 & 0.44 & 0.39 & 0.05 \\
In-plane O & 5  & 0.30 & 0.26 & 0.04 \\
\bottomrule
\end{tabular}
\label{tab:cebe_shift}
\end{table}

To assess the accuracy of the Z+1 approach for calculating O 1s core-electron binding energy (CEBE) shifts at SnO$_2$ (110) surfaces, we compare binding energy shifts of different surface O atoms (relative to an O atom in the middle of the slab) to results from full $\Delta$-SCF calculations, see Table~\ref{tab:cebe_shift}. We find that the CEBE shifts from the two methods are in very good agreement. The largest difference is 0.05 eV which is significantly smaller than the change of the binding energy shift for the different O atoms. We conclude that accurate CEBE shifts can be obtained from the Z+1 method.

\begin{figure}[H]
    \centering
    \includegraphics[width=0.95\textwidth]{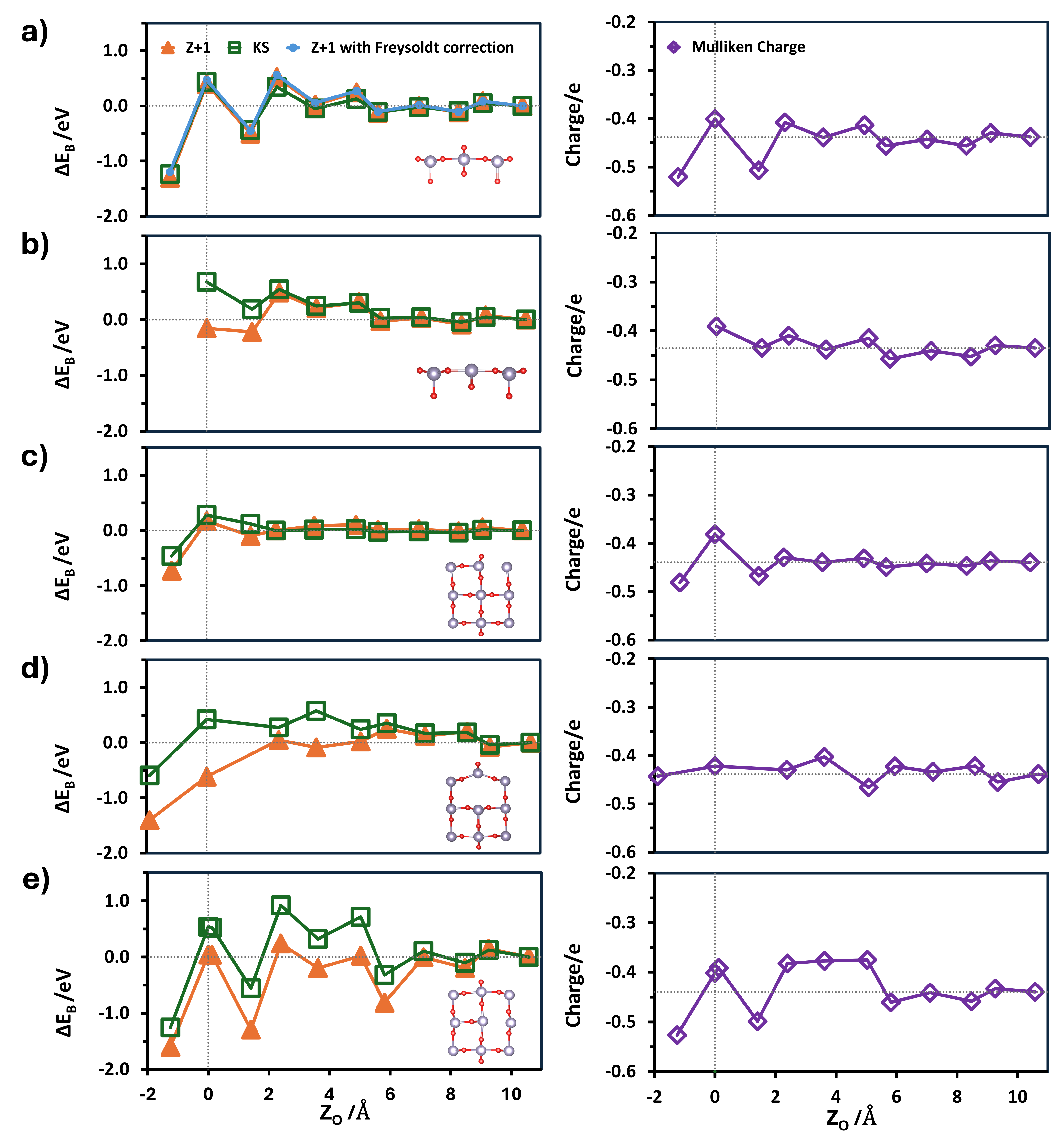}
    \caption{Left panels: O~1s core-electron binding energy (CEBE) shifts $\Delta E_B$ of SnO$_2$(110) surfaces from the Z+1 method and from Kohn-Sham (KS) eigenvalues as function of the depth $z_{\text{O}}$ of the O atom relative to the plane of in-plane O atoms (indicated by vertical dashed lines). Results are shown for the stoichiometric surface (a), the fully reduced surface (b), the surface with in-plane O vacancies (c), the surface with sub-bridging O vacancies (d), and the surface with sub-in-plane O vacancies (e). Right panels: Mulliken charges of the O atom in the various SnO$_2$ surfaces in units of the proton charge $e$. All CEBE shifts are referenced to the O atom in the center of the slab (indicated by the dashed horizontal line).}
    \label{fig:z+1_shift_all_UHV}
\end{figure}

Next, we determine CEBE shifts for different SnO$_2$ (110) surfaces without adsorbates using the Z+1 approach. The left panel of Fig.~\ref{fig:z+1_shift_all_UHV} shows the calculated O 1s CEBE shifts (relative to the O atom in the center of the slab which represents the bulk) as a function of the O atom's z-coordinate which is measured relative to the plane of in-plane O atoms. Shown are results from the Z+1 method obtained for a $5 \times 5$ supercell using a 7-layer slab. For the stoichiometric surface (panel a)) CEBE shifts with (blue dots) and without (orange triangles) the Freysoldt correction are shown. We find that for this supercell size the Freysoldt correction is always smaller than 0.1~eV. Because of the smallness of the correction, we do not include it for the other surfaces.

Inspecting the O 1s CEBE shifts for the stoichiometric surface, we observe that the bridging O atoms exhibit a very large ($\approx -1.30$~eV) negative shift, while the in-plane O atoms exhibit a positive shift of about 0.40~eV. The CEBE shifts continue to exhibit an oscillatory behaviour for the next O atoms and then approach the bulk value at about 6 \AA \; below the surface. To understand this behaviour of the CEBE shifts, we have calculated the Mulliken charges of the O atoms, see right panel of Fig.~\ref{fig:z+1_shift_all_UHV} a). We find that the Mulliken charge of the bridging O is more negative than that of the O atom in the center of the slab. It is therefore easier to remove an electron from the bridging O atom, explaining its smaller binding energy. In contrast, the Mulliken charge of the in-plane O atom is less negative than that of the bulk O atom giving rise to a positive value of the CEBE shift. The more negative Mullikan charge of the bridging O atoms is a consequence of their reduced (two-fold) coordination which results in highly polar bonds to its neighboring Sn atoms. In contrast, the in-plane O atoms have a threefold coordination and therefore less polar bonds.

\begin{figure}[H]
    \centering
    \includegraphics[width=0.95\textwidth]{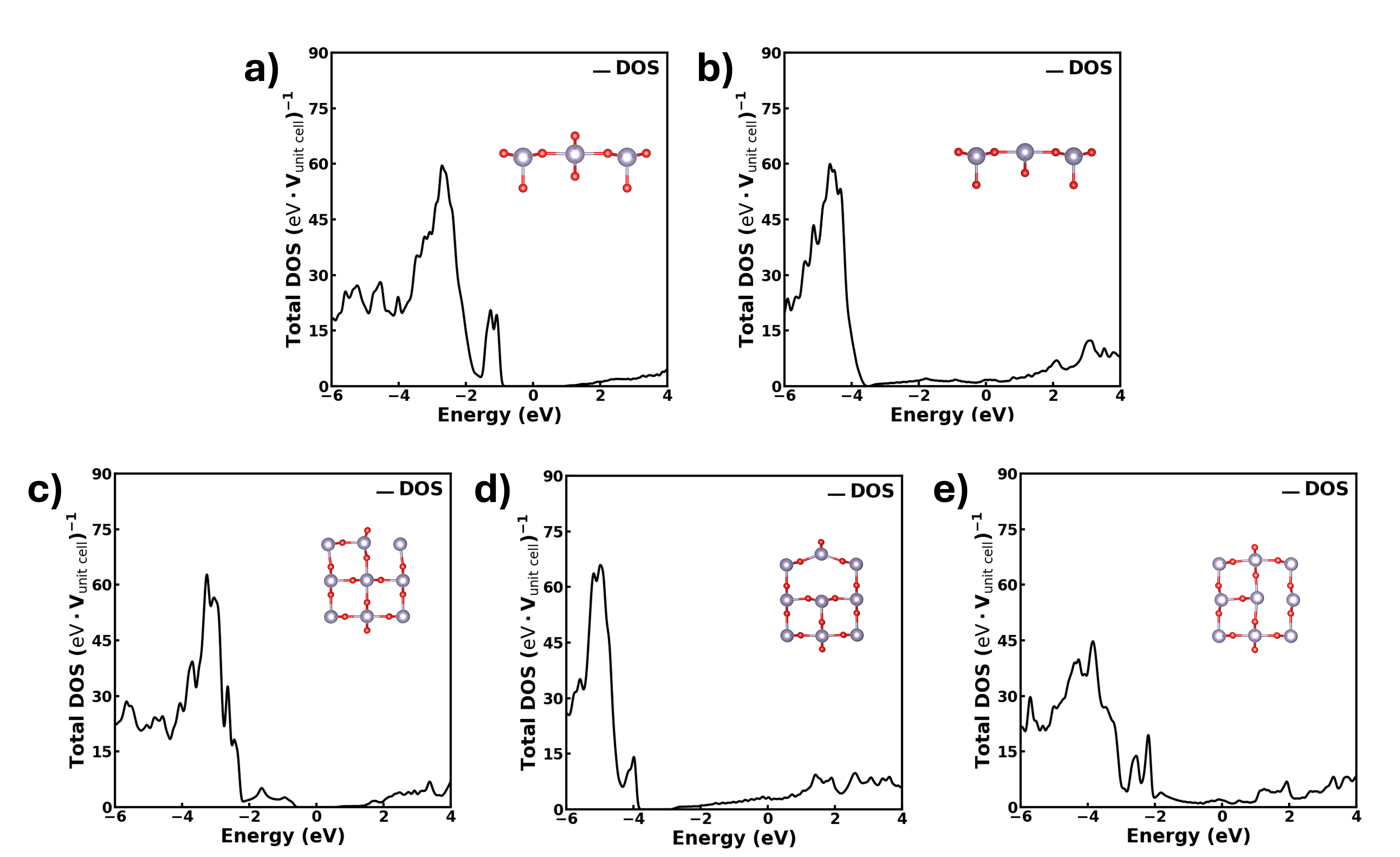}
    \caption{Density of states (DOS) of SnO$_2$(110) surfaces: a) stoichiometric surface, b) fully reduced surface, c) surface with in-plane O vacancies, d) surface with sub-bridging O vacancies, and (e) surface with sub-in-plane O vacancies. The Fermi level is set to 0 eV in all graphs. Insets show the corresponding relaxed surface structures. }
    \label{fig:DOS_UHV}
\end{figure}

For comparison, we also show the CEBE shifts obtained from ground-state Kohn-Sham energies, i.e. calculations without a core hole. Surprisingly, the CEBE shifts obtained from the Kohn-Sham energies are in very good agreement with the Z+1 results. This is because the stoichiometric surface is non-conducting (its density of states is shown in Fig.~\ref{fig:DOS_UHV} a)) and therefore the CEBE shifts are dominated by initial-state effects (such as variations of the atomic charges) which are captured by the Kohn-Sham eigenvalues. In contrast, for metallic surfaces, such as the fully reduced surface (discussed in the next paragraph), final-state effects also play an important role, resulting in a significant difference between the Z+1 results and the CEBE shifts obtained from the Kohn-Sham energies.

Next, we present O 1s CEBE shifts of the fully reduced (110) surface, see left panel of Fig.~\ref{fig:z+1_shift_all_UHV} b). In contrast to the stoichiometric surface, the O 1s CEBE shifts do not exhibit a strong oscillatory behaviour as a function of O atom depth: the CEBE shift of the in-plane O atom is -0.15 eV relative to the bulk O atom, and the CEBE shift of the next O atom is -0.22 eV. The next O atom has a positive CEBE shift of 0.49 eV, and then the CEBE shifts approach the bulk value.

Interestingly, the CEBE shifts do not follow the Mulliken charges for the fully reduced surface, shown in the right panel of Fig.~\ref{fig:z+1_shift_all_UHV} b). In fact, the Mulliken charge of the in-plane O atoms is less negative than the charge of the bulk atom, and one would therefore expect that its CEBE shift should be positive. As noted above, however, the fully reduced surface is metallic, see Fig.~\ref{fig:DOS_UHV} b), and the free carriers screen the core hole giving rise to pronounced final-state effects.

To understand the role of final-state effects, we compare the Z+1 CEBE shifts to the result obtained from Kohn-Sham energies. The binding energies obtained from the Kohn-Sham eigenvalues again exhibit a similar behaviour as the Mulliken charges. We observe that the Kohn-Sham results agree with the Z+1 results for the O atoms far from the surface. However, for the topmost two O atoms, qualitatively different results are obtained: based on the Kohn-Sham eigenvalues, a positive CEBE shift is predicted for these O atoms, while the Z+1 method predicts a negative shift. This difference is caused by final state effects, which are not captured by the Kohn-Sham eigenvalues. The final-state effects stabilize the core hole, thus reducing the CEBE. As the free charge carriers are localized at the surface (the bulk remains semiconducting), only the surface O atoms exhibit a reduction of the CEBE shift. For the in-plane O atom, we find the shift due to final-state effects (defined as the difference between the Z+1 results and the Kohn-Sham result for the CEBE shift) to be 0.83~eV. 

We also consider surfaces with various other types of O vacancies. Fig.~\ref{fig:slab_model_UHV} c) shows CEBE shifts for a surface with in-plane O vacancies. Similar to the stoichiometric surface, the binding energies exhibit an oscillatory behaviour but with a significantly reduced amplitude: the CEBE shift of the bridging O atom is only -0.7 eV, while that of the remaining in-plane O atom is 0.2 eV. This surface is non-metallic (see Fig.~\ref{fig:DOS_UHV} c)) and therefore the Z+1 results are in close agreement with those obtained from the Kohn-Sham energies. The smaller CEBE shifts are a consequence of the small variations of the Mulliken charges: the vacancy attracts electrons from nearby O atoms (in particular, from bridging and sub-bridging O atoms) whose Mulliken charges become less negative.

Removal of sub-bridging O atoms produces a metallic surface, see Fig.~\ref{fig:DOS_UHV} d). The associated final-state effects give rise to significant differences between Z+1 and Kohn-Sham CEBE shifts: except for the bridging O atom, the Kohn-Sham method predicts positive CEBE shifts for the O atoms near the surface, while the Z+1 approach predicts negative CEBE shifts (-1.4 eV for the bridging O atom and -0.6 eV for the in-plane O), see Fig.~\ref{fig:slab_model_UHV} d). Finally, the surface with sub-in-plane O vacancies is also metallic. For this system, large negative CEBE shifts are predicted for the bridging and sub-briding O atoms, see Fig.~\ref{fig:slab_model_UHV} e). These are caused by a combination of large negative atomic charges and final-state screening effects.

\begin{figure}[H]
    \centering
    \includegraphics[width=0.95\textwidth]{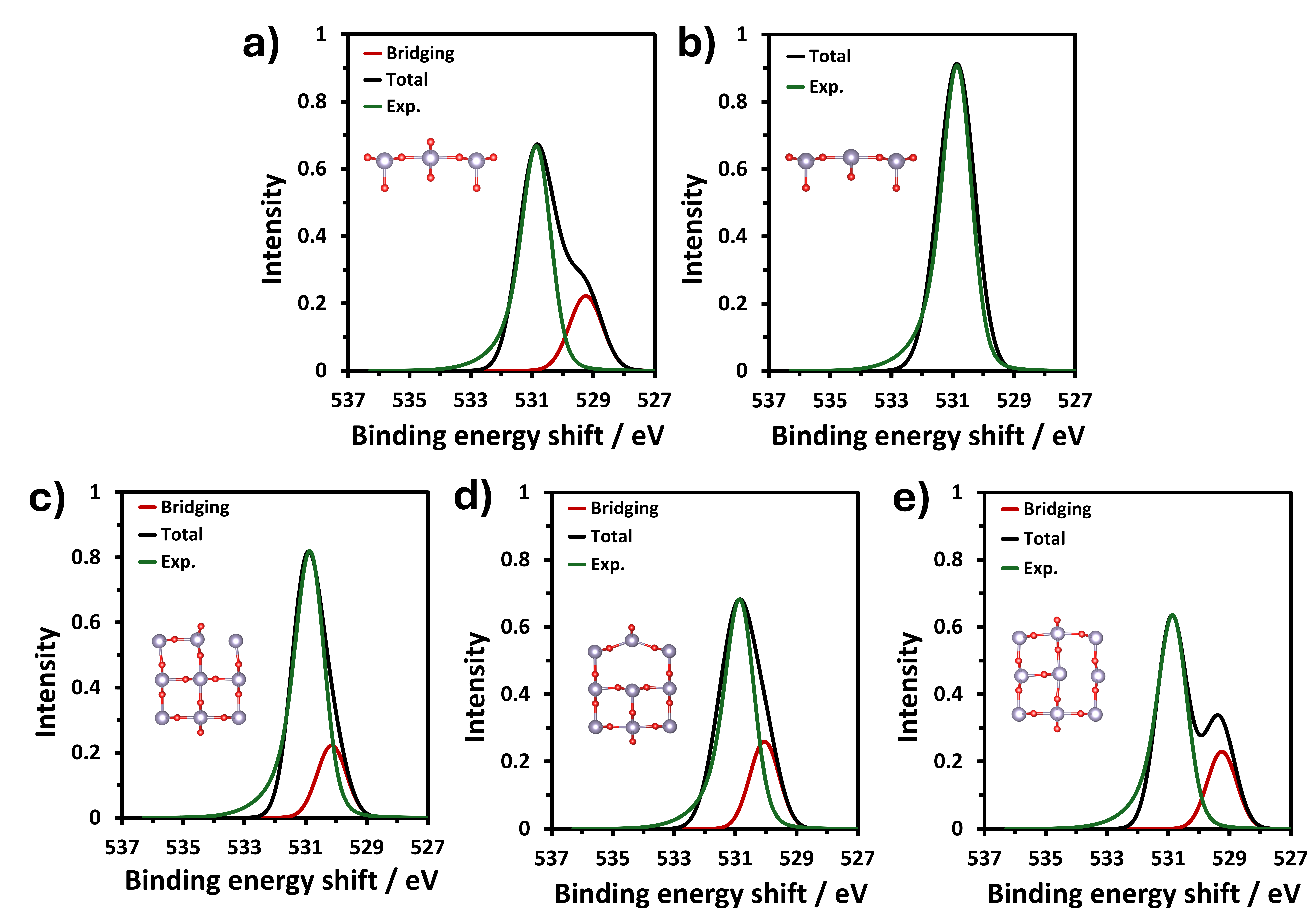}
    \caption{Calculated O~1s XPS spectra (using the Z+1 method)  of five SnO$_2$(110) surfaces: a) stoichiometric surface, b) fully reduced surface, c) surface with in-plane O vacancies, d) surface with sub-bridging O vacancies, and e) surface with sub-in-plane O vacancies. Also shown is experimental O 1s spectrum from the paper of Kucharski and coworkers~\cite{kucharski2022direct} obtained for reduced surfaces.}
    \label{fig:xps_comparison_UHV}
\end{figure}

To enable a direct comparison with measured O 1s XPS spectra, we present the simulated spectra of the different SnO$_2$ surfaces in Fig.~\ref{fig:xps_comparison_UHV}. Also shown are the experimental O 1s spectra obtained by Kucharski et al.~\cite{kucharski2022direct} of reduced surfaces before exposure to O$_2$. The experimental spectrum was recorded using a photon energy of 895 eV and experiments were performed at 300 $^\circ$C. The measured O 1s spectrum is highly symmetric in agreement with the calculated spectrum of the fully reduced surface shown in panel b). In contrast, the calculated spectrum of the stoichiometric surface exhibits an additional peak due to bridging O atoms at low binding energies which is not observed in the experiment. Similar peaks at low-binding energies are found in the surface with in-plane, sub-bridging and sub-in-plane O vacancies. We conclude that the surface studied by Kucharski and coworkers before exposure to oxygen resembles most closely the fully reduced surface. In the following, we therefore study the adsorption of molecules at the fully reduced surface.

\begin{figure}[H]
    \centering
    \includegraphics[width=0.85\textwidth]{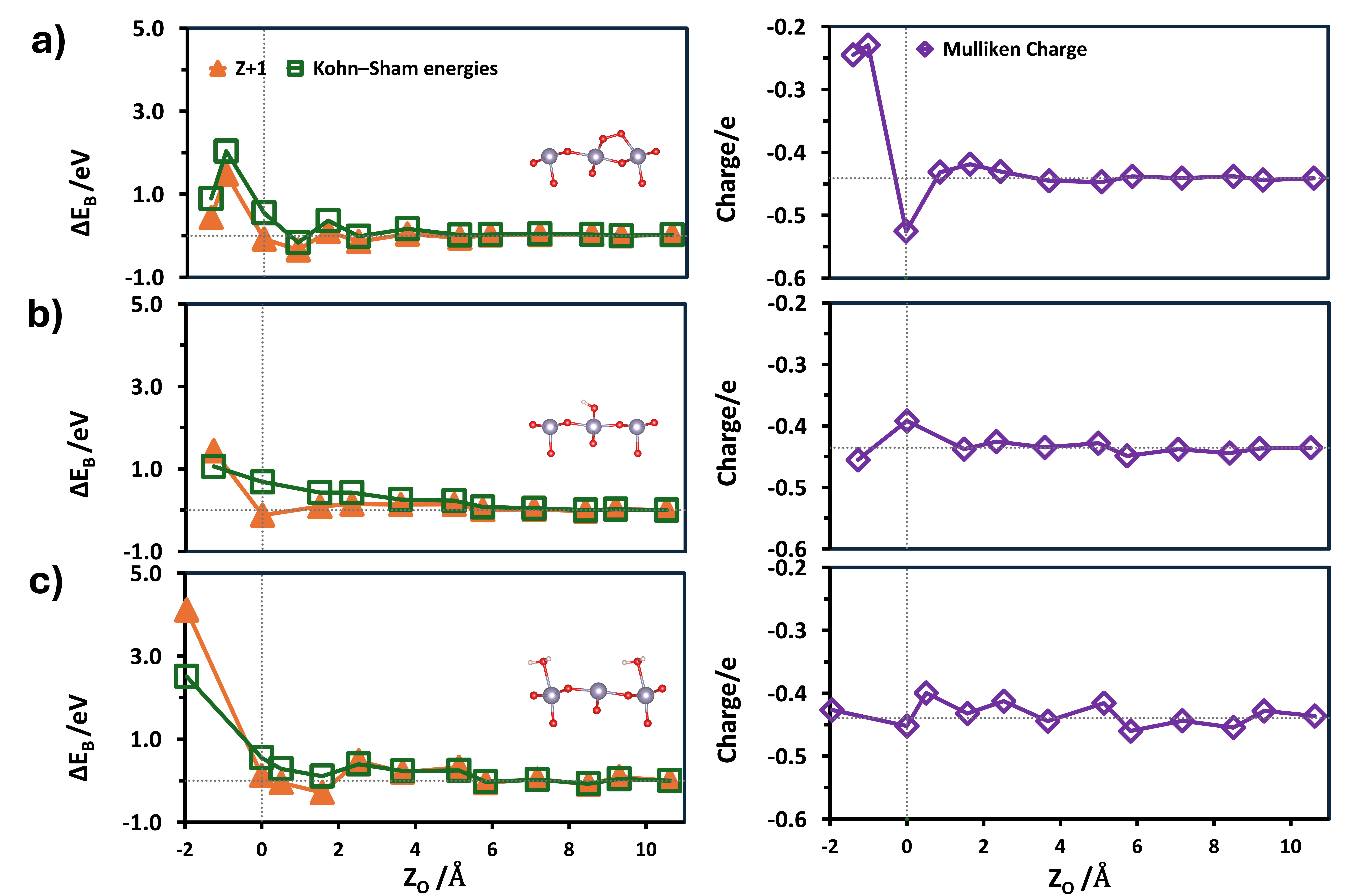}
    \caption{Left panels: O~1s core-electron binding energy (CEBE) shifts $\Delta E_B$ of SnO$_2$(110) surfaces from the Z+1 method and from Kohn-Sham (KS) eigenvalues as function of the depth $z_{\text{O}}$ of the O atom relative to the plane of in-plane O atoms (indicated by vertical dashed lines). Results are shown for the fully reduced surface with adsorbed O$_2$ molecules (a), the fully reduced surface with an adsorbed OH (b), and the fully reduced surface with H$_2$O molecules (c). Right panels: Mulliken charges of the O atom in the various SnO$_2$ surfaces in units of the proton charge $e$. All CEBE shifts are referenced to the O atom in the center of the slab (indicated by the dashed horizontal line).}
    \label{fig:z+1_shift_all_O2}
\end{figure}

\begin{figure}[H]
    \centering
    \includegraphics[width=0.95\textwidth]{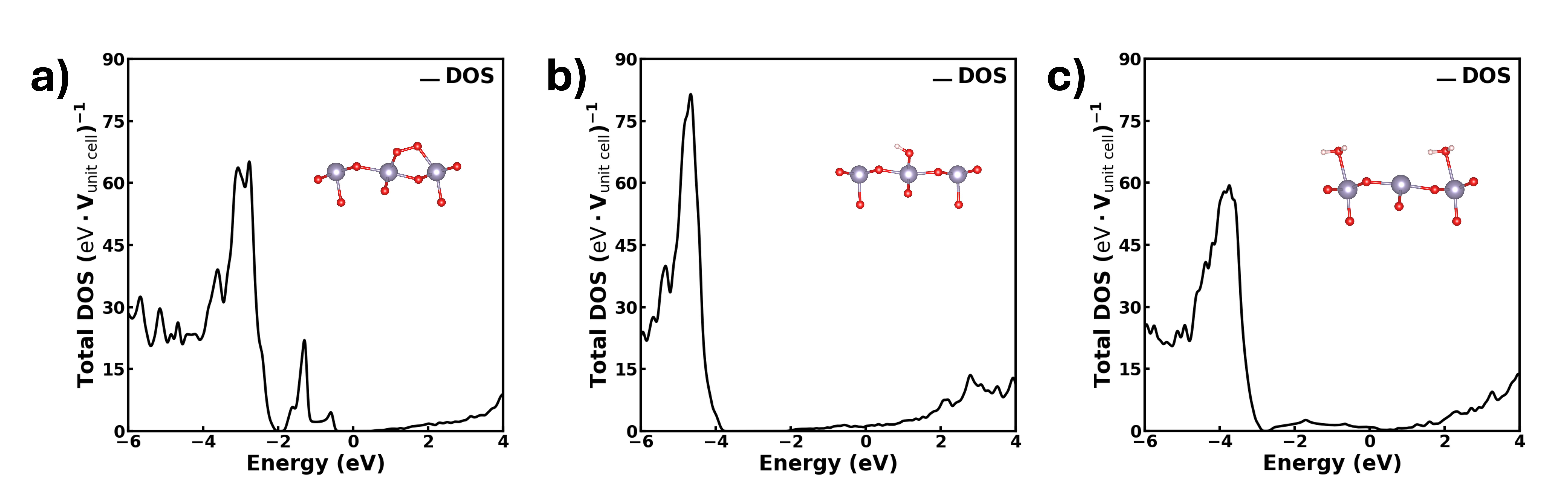}
    \caption{Density of states (DOS) of SnO$_2$(110) surfaces: a) fully reduced surface with adsorbed O$_2$ molecules, b) fully reduced surface with an adsorbed OH, and c) fully reduced surface with H$_2$O molecules. The Fermi level is set to 0 eV in all graphs. Insets show the corresponding relaxed surface structures. }
    \label{fig:DOS_O2}
\end{figure}

First, we consider the fully reduced SnO$_2$ surfaces with adsorbed O$_2$ molecules (Fig.~\ref{fig:z+1_shift_all_O2} a)), adsorbed OH (Fig.~\ref{fig:z+1_shift_all_O2} b)), and adsorbed H$_2$O molecules (Fig.~\ref{fig:z+1_shift_all_O2} c)). For the fully reduced surface with adsorbed O$_2$, the Z+1 method predicts positive CEBE shifts for the two O atoms of the O$_2$ molecule. Specifically, the shift for the outermost O atom is 0.44~eV and for the O atom in the bridging O position the shift is 1.49~eV. The in-plane O atom and all other O atoms have smaller CEBE shifts. Comparing the Z+1 results to the Kohn-Sham results, we find that both methods predict similar CEBE shifts (with the Z+1 shifts being a few tenths of an 0.46~eV smaller than the Kohn-Sham results for the outmost O atoms). This similarity can be attributed to the non-metallic character of the surface, see Fig.~\ref{fig:DOS_O2} a). We find a strong correlation between the Z+1 results for the CEBE shifts and the Mulliken charges. In particular, the Mulliken charges of the outermost two O atoms are significantly less negative than the charge of a bulk O atom: they are about half of the charge on a bulk O atom. This suggests that the O$_2$ molecule carries a charge of $-2$e, i.e. the molecule accepts two electrons from the surface upon adsorption. 

For the hydroxylated surface, we find a large positive CEBE shift (approx. 1.44~eV) for the O atom of the OH group, while the in-plane O atom exhibits a small negative shift. The next O atoms have very small positive shifts relative to a bulk O atom. With the exception of the O atom of the OH group, we find that the O atoms near the surface exhibit a smaller CEBE shift in the Z+1 results compared to the Kohn-Sham results. This difference can again be attributed to final-state effects as this surface is metallic (see Fig.~\ref{fig:DOS_O2} b). To understand the large positive CEBE shift of the O atom of the OH group, we note that this system can also be viewed as a stoichiometric surface in which the bridging O atoms are chemically bonded to H atoms. Therefore, these O atoms have to share their electrons with the adsorbed H atom which reduces their negative charge and therefore produces more positive CEBE shifts. Inspecting the Mulliken charge of the outermost O atoms, see right panel of Fig.~\ref{fig:z+1_shift_all_O2} b), we indeed find that it is less negative than that of the bridging O atom of the stoichiometric surface. Surprisingly, however, the Mulliken charge is still more negative than in the bulk. We attribute this to the inability of the Mulliken charge to accurately capture the less ionic bond between the O and H atoms of the hydroxyl group.

For the fully reduced surface with adsorbed H$_2$O, we observe a large positive CEBE shift (approx. 4.1 eV) for the adsorbate oxygen atom, see Fig.~\ref{fig:z+1_shift_all_O2} c). Similar to the case of OH, this large CEBE shift cannot be explained by the Mulliken charges: the Mulliken charge of the adsorbate O atom is close to the SnO$_2$ bulk result. Again, we believe that this is a consequence of the inability of the Mulliken charge to accurate describe the charge distribution of the covalent OH bond.

\begin{figure}[H]
    \centering
    \includegraphics[width=0.95\textwidth]{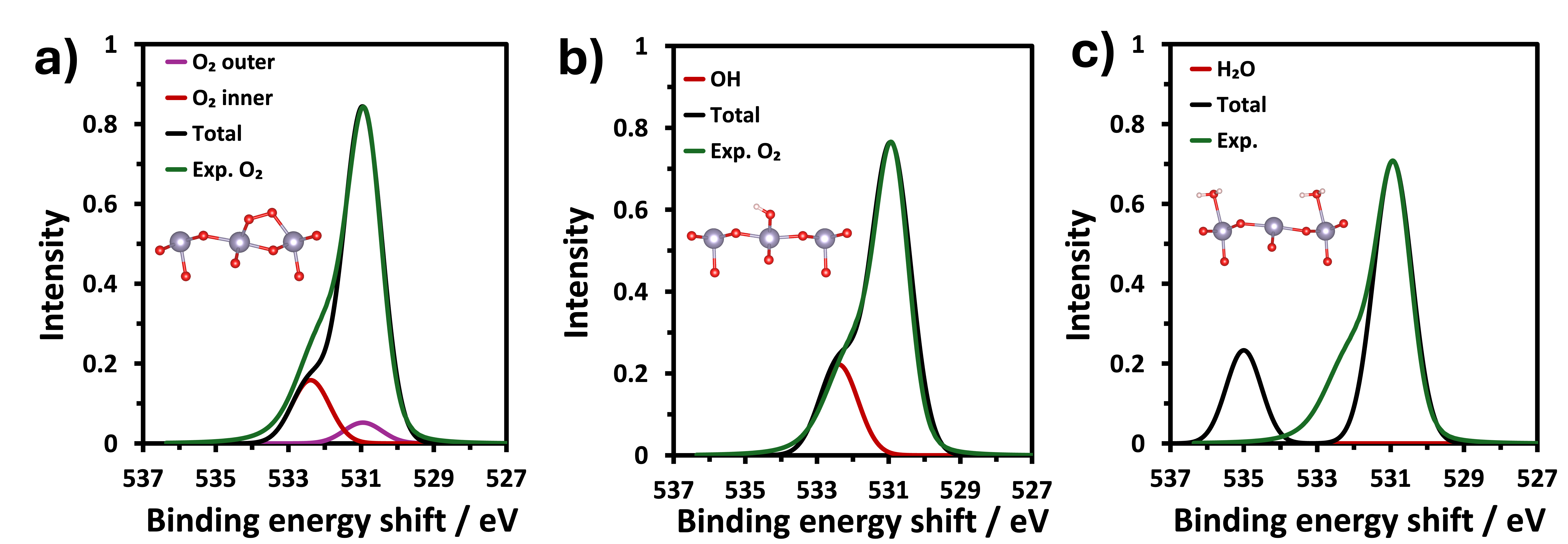}
    \caption{Calculated O~1s XPS spectra (using the Z+1 method)  of three SnO$_2$(110) surfaces: a) fully reduced surface with adsorbed O$_2$ molecules, b) fully reduced surface with an adsorbed OH, and c) fully reduced surface with H$_2$O molecules. Also shown is the experimental O 1s spectrum from the paper of Kucharski and coworkers~\cite{kucharski2022direct} obtained for surfaces exposed to oxygen gas.}
    \label{fig:xps_comparison_O2}
\end{figure}

After exposure to oxygen gas the experimental spectrum exhibits a high-energy shoulder as shown in Fig.~\ref{fig:xps_comparison_O2}. The calculated spectra of the fully reduced surface with adsorbed O$_2$ and OH also exhibit such a high-energy feature. This feature is caused by the adsorbate oxygen atoms. In contrast, the adsorbate peak of the surface with adsorbed water molecules is at significantly higher binding energies. Kucharski et al. exclude the possibility of adsorbed OH on their surfaces and assign the high-energy shoulder in their spectrum to adsorbed oxygen molecules~\cite{kucharski2022direct}. This interpretation is consistent with our calculations. We note that similar spectra were obtained by Voroktha and coworkers~\cite{VOROKHTA2018284}. However, these authors assign the high-energy peak to adsorbed OH groups rather than O$_2$ molecules. Our calculations show that this interpretation of the experimental spectra is also possible as the CEBE shifts adsorbed OH and adsorbed O$_2$ are almost the same. Recently, significant concern has been raised regarding the characterization of peaks in XPS O1s spectra as arising from oxygen vacancies~\cite{IDRISS2021} in many current published experimental materials characterization of oxides, particularly in the context of air-exposed samples~\cite{Easton2025}. Ironically, our calculations show that the peak typically labelled as ‘oxygen vacancies’ instead arises from ‘additional oxygen’ in the form of OH or bridging oxygen species formed from interaction with a bridging oxygen vacancy, and the results here are also consistent with the suggestion of Liu et al.~\cite{Liu2023} and Li et al.~\cite{li2023} for an equivalent spectroscopic signature in ZnO to arise from adsorbed water or OH.

\section{Conclusions}

We have presented a first-principles approach for calculating the XPS spectrum of complex oxide surfaces: first, core-electron binding energies of atoms near the surface are determined using the Z+1 method and then the spectrum is constructed by adding the contributions of all atoms (taking into account the photo-electron mean-free path). This method is applied to various SnO$_2$ (110) surfaces (stoichiometric surface, surfaces with different types of vacancies as well as fully reduced surfaces with OH groups and adsorbed oxygen molecules) and the predicted spectra are compared to recent measurements. We find that the calculated O 1s spectrum for the fully reduced surface is highly symmetric and in very good agreement with experiment. In contrast the calculated spectra for the surfaces with OH and O$_2$ adsorbates exhibit additional features at higher binding energies. This finding is in agreement with measurements taken on surfaces that are exposed to oxygen gas. We conclude that the approach presented in this work is powerful tool to support the analysis of experimental XPS spectra and can provide much-needed insights into the chemical environments present at complex oxide surfaces.

\begin{acknowledgement}

\end{acknowledgement}
JL acknowledges funding from the Royal Society through a Royal Society University Research Fellowship URF/R/191004. We acknowledge computational resources from ARCHER2 UK National Computing Service which was granted via HPC-CONEXS, the UK High-End Computing Consortium (EPSRC grant no. EP/X035514/1). CB acknowledges support from the Engineering and Physical Sciences Research Council (EP/R512400/1).  This project has received funding from the European Union's Horizon Europe research and innovation programme under grant agreement no. 101131173. JMK acknowledges support from the European Union’s Horizon Europe project EXANST (contract 101159716), from the Estonian Ministry of Education and Research grant number TK210 and from the Estonian Research Council grant number PSG1037.
%JL: CDT? BETTERXPS?

\bibliography{achemso-demo}

\end{document}